\documentclass[reprint,prd,twocolumn]{revtex4-1}

\usepackage{graphicx}
\graphicspath{{./graphics/}}

\usepackage{amsmath}
\usepackage{amssymb}
\usepackage{url}
\usepackage{float}
\usepackage{tikz}

\ifcsname italic \endcsname%
\else%
  
\fi
\newcommand{\ratio}{{$\sigma_{e^+p}/\sigma_{e^-p}$} }
\newcommand{\dg}{{$^\circ$} }
\newcommand{\ep}{{$e^- p$} }
\newcommand{\pp}{{$e^+ p$} }
\newcommand{\pmp}{{$e^\pm p$} }
\newcommand{\rtg}{{$R_{2\gamma}$} }

\newcommand{\td}[2]{\frac{\mathrm{d} #1}{\mathrm{d} #2}}

\makeatletter
\g@addto@macro\bfseries{\boldmath}
\makeatother

\begin{document}

\title{Measurement of the Two-Photon Exchange Contribution to Elastic Lepton-Proton Scattering at the OLYMPUS Experiment}

\author{Brian S. Henderson}
\email{bhender1@mit.edu}
\affiliation{Massachusetts Institute of Technology, Laboratory for Nuclear Science, Cambridge, MA 02139, USA}
\collaboration{for the OLYMPUS Collaboration}
\noaffiliation

\date{September 27, 2016}

\begin{abstract}
Measurements of the ratio of the proton elastic form factors ($\mu_pG_E/G_M$)
using Rosenbluth separation and those using polarization-based techniques show a
strong discrepancy, which increases as a function of $Q^2$.  The contribution of
hard two-photon exchange (TPE) to \ep scattering, which is neglected in the standard treatments of
elastic \ep scattering, is the most widely-accepted hypothesis for the explanation of this discrepancy.
While calculations of the hard TPE contribution are highly model dependent, the effect may be quantified experimentally
by precisely measuring the ratio of the positron-proton and electron-proton elastic scattering cross sections.
The OLYMPUS experiment collected approximately 4 fb$^{-1}$ of \pp and \ep scattering data at the DORIS storage ring at DESY in 2012, with the
goal of measuring the elastic \ratio ratio over the kinematic range $(0.4 \leq \epsilon \leq 0.9)$, $(0.6 \leq Q^2 \leq 2.2)$ GeV$^2/c^2$ at a
fixed lepton beam energy of 2.01 GeV.  Initial results from OLYMPUS were presented, and subsequently results on \rtg = \ratio from the experiment
have been publicly released and are in preparation for publication \cite{prlsubmit}.
\end{abstract}

\maketitle{}

\section{The Form Factor Discrepancy and Two-Photon Exchange}

Measurements of the proton's elastic form factors over several decades using
Rosenbluth separation \cite{PhysRev.79.615} with inclusive \ep scattering experiments persistently showed results
for the ratio of the form factors ($\frac{\mu_pG_E}{G_M}$) consistent with unity up to four-momentum transfers of
up to $\sim$10~(GeV/$c$)$^2$ \cite{ff3,ff4,ff8,ff9}.
Extracting the form factors in this fashion, however, requires careful consideration of radiative corrections to
the measured cross section and detailed treatment of systematic uncertainties that affect extraction of the 
cross sections at different kinematics.  The advent of polarized beams and targets in the 1990s provided a new method
of measuring the ratio of the form factors using the relative cross sections for \ep scattering for various
combinations of the incoming and outgoing particle polarizations.  For example, for longitudinally polarized
electrons scattered from unpolarized protons the cross sections for the outgoing proton to have longitudinal
or transverse polarization are given by
\begin{equation}
 \td{\sigma^{(L)}}{\Omega} = h\sigma_\text{Mott} \frac{E+E'}{m_p}\sqrt{\frac{1}{1+\frac{4m_p^2}{Q^2}}} \tan^2\left(\frac{\theta}{2}\right) G_M^2,
\end{equation}
\begin{equation}
 \td{\sigma^{(T)}}{\Omega} = 2h\sigma_\text{Mott}\sqrt{\frac{1}{1+\frac{4m_p^2}{Q^2}}} \tan\left(\frac{\theta}{2}\right) G_E G_M,
\end{equation}
respectively \cite{Arrington:2011dn}.  The ratio of the form factors is then proportional to the ratio of the measured
proton polarizations.  While polarization-based methods do not provide a way of measuring the form factors individually, the extraction
of a ratio of cross sections provides a means of canceling various systematic uncertainties and contributions from
radiative corrections.

\begin{figure}[thb!]
\centerline{\includegraphics[width=\columnwidth]{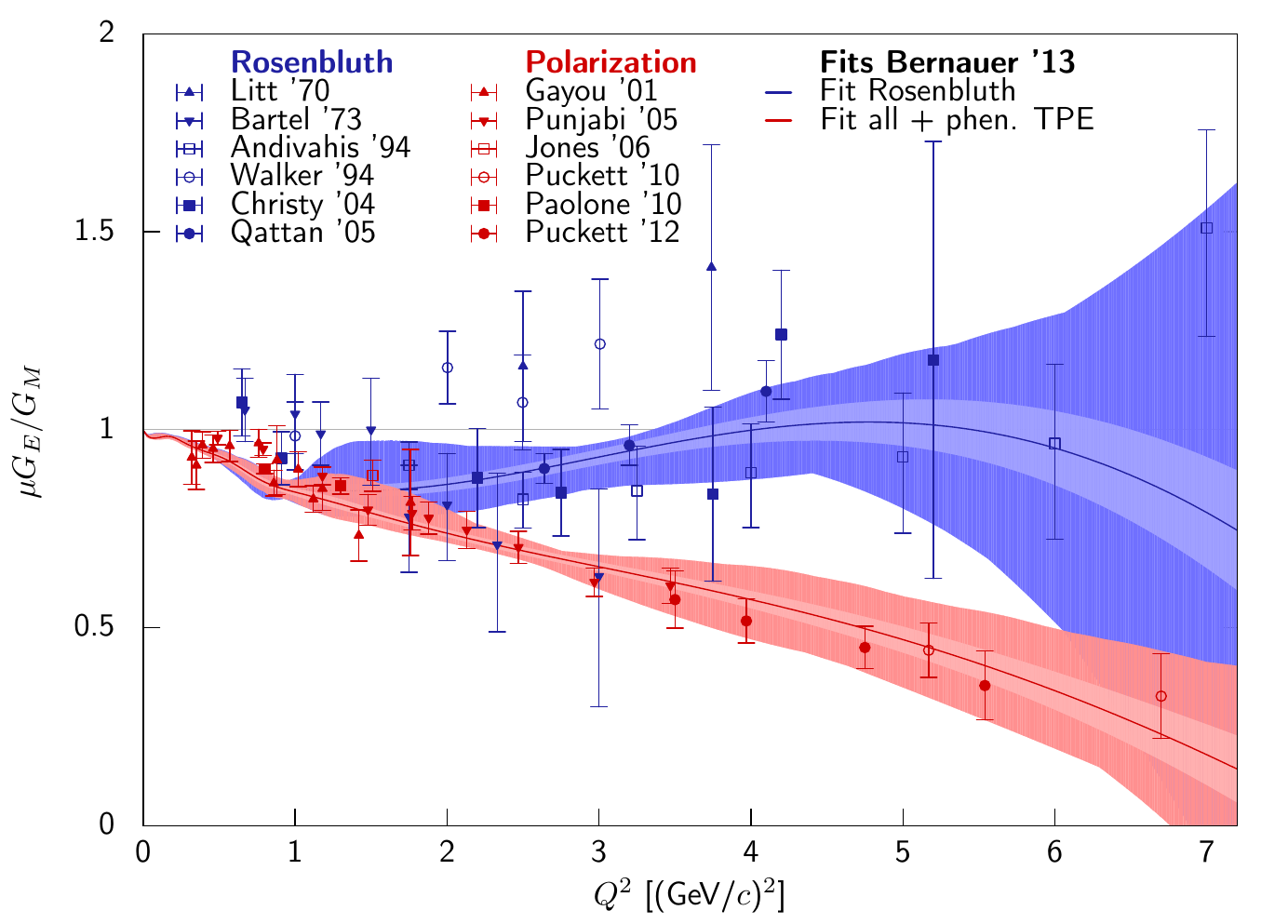}}
\caption{Selection of experimental results on the ratio $\frac{\mu_pG_E}{G_M}$ along with phenomenological fits to the form factor data, illustrating the discrepancy
between experiments using Rosenbluth separation and polarization-based methods.  
(Rosenbluth separation data: \cite{ff3,ff4,ff8,ff9,ff10,ff11}, polarization data: \cite{pol11,pol3,pol12,pol7,pol9,pol13}, phenomenological fit: \cite{BerFFPhysRevC.90.015206}).}
\label{fig:disc}
\end{figure}

As shown in Figure \ref{fig:disc}, extractions of $\frac{\mu_pG_E}{G_M}$ from polarization-based experiments \cite{pol11,pol3,pol12,pol7,pol9,pol13}
showed a decreasing trend with $Q^2$ in sharp contrast to the Rosenbluth separation results.  Modern Rosenbluth separation experiments using
exclusive \ep event reconstruction \cite{ff10,ff11} as well as reanalyses of past experiments \cite{PhysRevC.69.022201,PhysRevC.68.034325} confirmed
the discrepancy.  Since the form factors cannot be calculated directly from first principles, this discrepancy represents an important issue in
hadronic physics that calls into question knowledge of the proton's most fundamental properties.

\begin{figure}[thb!]
\centerline{\includegraphics[width=\columnwidth]{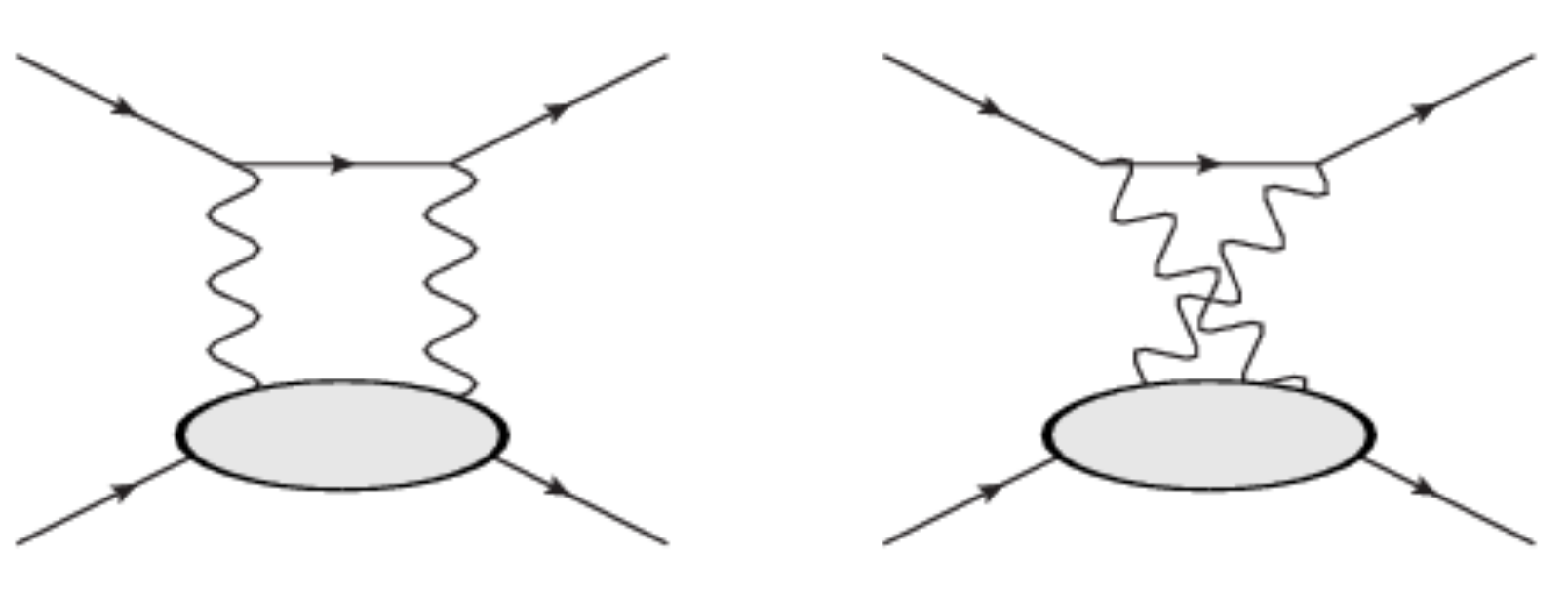}}
\caption{The ``box'' and ``crossed box'' two-photon exchange diagrams that contribute to the overall amplitude for elastic \ep scattering.  The nature of the
intermediate hadronic state connecting the incoming and outgoing protons makes calculations of these diagrams difficult and model-dependent.}
\label{fig:tpediag}
\end{figure}

The most widely-accepted hypothesis for the explanation of the discrepancy is that contributions from hard two-photon exchange (TPE), which are
neglected in the standard radiative corrections applied to Rosenbluth scattering data \cite{PhysRev.122.1898,MoRevModPhys.41.205,MaximonPhysRevC.62.054320}, could
alter the results of the Rosenbluth separation measurements to bring them into agreement with those from polarization-based experiments.  Contributions
from the TPE diagrams, shown in Figure \ref{fig:tpediag}, to the elastic \ep cross section are difficult to calculate due to the lack of model-independent
methods for the description of the intermediate hadronic state.  A number of theoretical and phenomenological calculations for the TPE contribution to elastic \ep scattering and its effect
on the form factor ratio were put forward \cite{Blunden:2003sp,Guichon:2003qm,REKALO2004322,Chen:2004tw,Afanasev:2005mp,Blunden:2005ew, Kondratyuk:2005kk,Borisyuk:2006fh,TomasiGustafsson:2009pw},
but the model dependence and the spread among predictions demanded an experimental determination of the TPE contribution.

\section{Experimental Signature of Two-Photon Exchange}

The contribution to the elastic \ep cross section from the TPE diagrams is predicted to be an effect of at most several percent at the accessible
kinematics \cite{BerFFPhysRevC.90.015206,Chen:2007ac, Guttmann:2010au, Blunden:2003sp,Chen:2004tw,Afanasev:2005mp,Blunden:2005ew, Kondratyuk:2005kk, Borisyuk:2006fh,TomasiGustafsson:2009pw},
and thus precise experimental techniques are required to measure deviations from the standard radiative cross section prescriptions.  Such a precise measurement may be
made by taking advantage of the fact that the leading-order contribution from TPE to the total elastic lepton-proton scattering matrix element arises from the interference
term between the one- and two-photon exchange diagrams, which carries an overall factor of the sign of the lepton charge.  Thus, letting $\mathcal{M}_\gamma$ and $\mathcal{M}_{\gamma\gamma}$
represent the one- and two-photon exchange matrix element contributions respectively, a measurement of the ratio of the \ep and \pp elastic cross sections provides a means
of directly accessing the magnitude of the $\mathcal{M}_{\gamma\gamma}$ contribution:
  \begin{equation}
   R_{2\gamma}\left(\epsilon,Q^2\right) = \frac{\sigma_{e^+}}{\sigma_{e^-}} \sim 1 + 4\alpha \frac{\mathcal{M}_{\gamma\gamma}}{\mathcal{M}_{\gamma}}.
  \end{equation}
Such an approach offers the advantages of canceling systematic uncertainties via measurement of a ratio of cross sections rather than absolute
cross section determination, although careful consideration must be given to effects such as $\alpha^3$ radiative corrections and detector acceptance
effects that can also induce differences between the \ep and \pp cross sections.  Additionally, precise measurement of the relative integrated
luminosity of \ep and \pp data taken is critical to the determination of the value of \ratio.

\section{The OLYMPUS Experiment}

\begin{figure}[thb!]
\centerline{\includegraphics[width=\columnwidth]{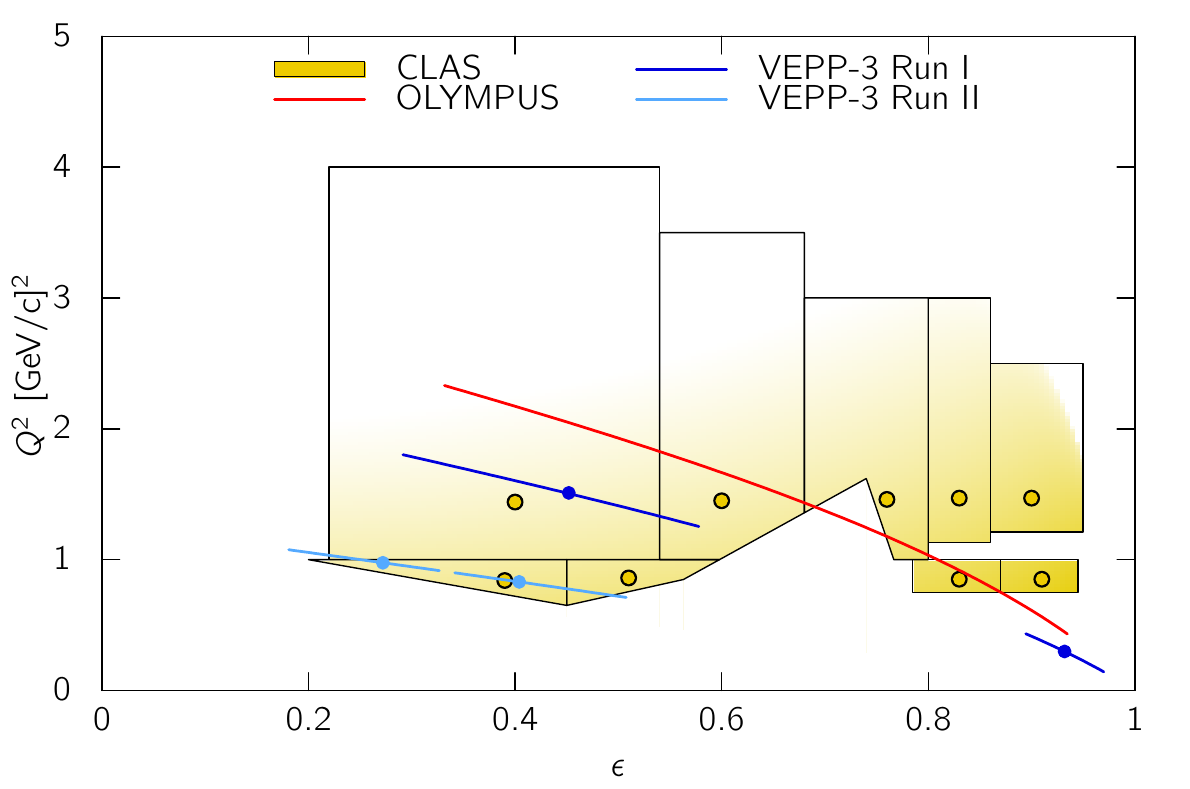}}
\caption{Kinematic reaches of the three modern experiments that have measured \ratio: OLYMPUS \cite{Milner:2014,prlsubmit}, CLAS \cite{PhysRevLett.114.062003,clasprl},
and VEPP-3 \cite{vepp3PhysRevLett.114.062005}.  Note that OLYMPUS and VEPP-3 operated at fixed beam energies (resulting in one-to-one correspondence between $\epsilon$
and $Q^2$) while CLAS operated with a variable energy tertiary beam requiring integration over 2D bins in the kinematic space.  (Figure reproduced from \cite{schmidt}.)}
\label{fig:reach}
\end{figure}

\begin{figure}[thb!]
\centerline{\includegraphics[width=\columnwidth]{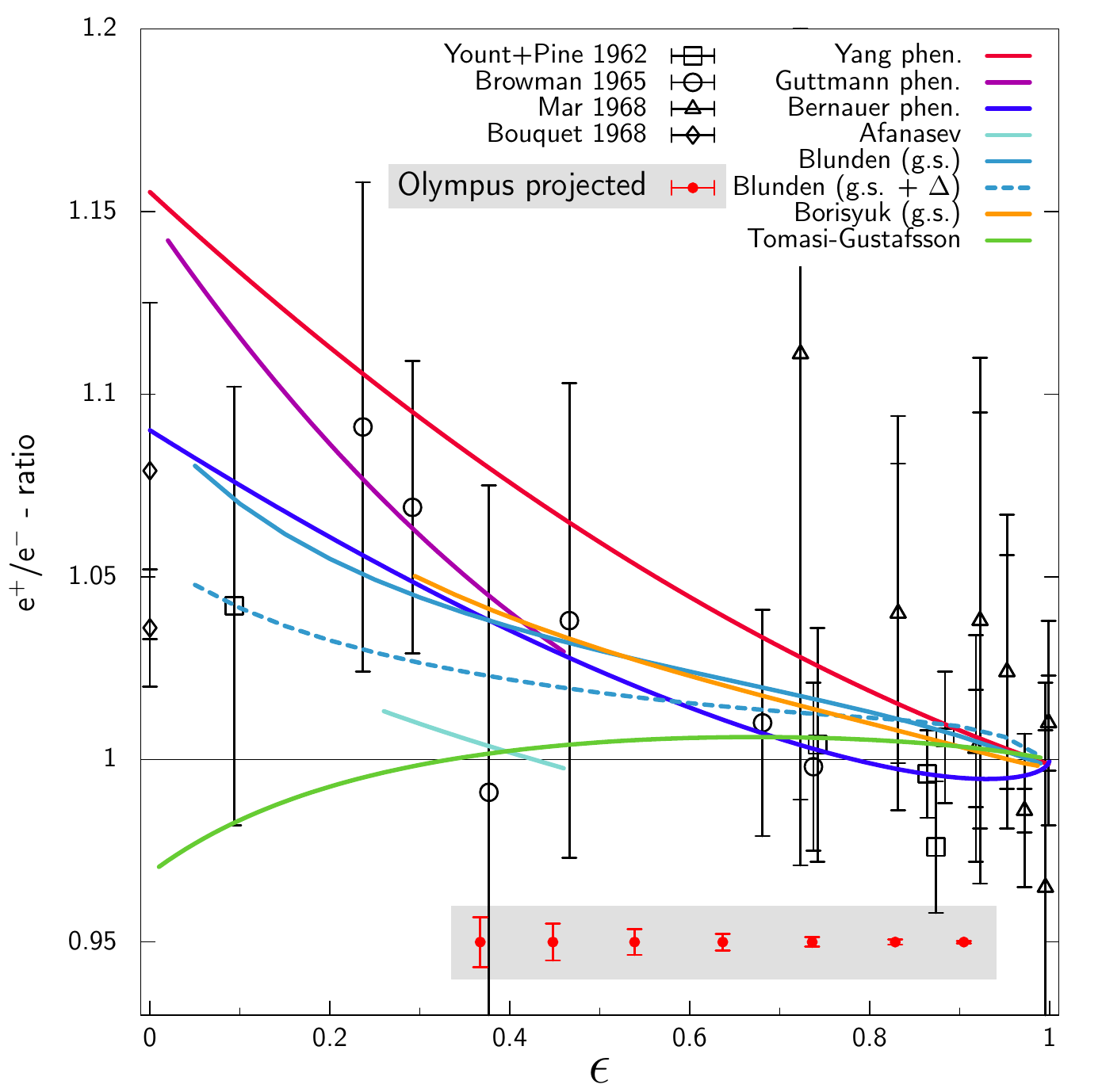}}
\caption{Projected uncertainty of the OLYMPUS \ratio measurement in comparison with existing data from the 1960s \cite{Yount:1962aa,Browman:1965zz,Bouquet:1968aa,Mar:1968qd}
and various phenomenological \cite{BerFFPhysRevC.90.015206,Chen:2007ac, Guttmann:2010au}
and theoretical \cite{Blunden:2003sp,Chen:2004tw,Afanasev:2005mp,Blunden:2005ew, Kondratyuk:2005kk, Borisyuk:2006fh,TomasiGustafsson:2009pw} predictions.}
\label{fig:proj}
\end{figure}

The OLYMPUS experiment was designed to measure \ratio using exclusively reconstructed \pmp events over the kinematic range $(0.4 \leq \epsilon \leq 0.9)$, $(0.6 \leq Q^2 \leq 2.2)$ GeV$^2/c^2$, 
corresponding to a fixed lepton beam energy of 2.01 GeV, with an uncertainty of $\lesssim$1\% over the full range.  The kinematic reaches of OLYMPUS and the other
two TPE experiments (CLAS \cite{PhysRevLett.114.062003,clasprl} and VEPP-3 \cite{vepp3PhysRevLett.114.062005}) are shown in Figure \ref{fig:reach}, while Figure
\ref{fig:proj} shows the projected uncertainties of the OLYMPUS measurement of \ratio in comparison to existing data from the 1960s and various theoretical
and phenomenological predictions.  While the form factor ratio discrepancy is most obvious at higher values of $Q^2$ than are accessible by the three
experiments, percent-level determination of \ratio at these kinematics (where the measurement of elastic scattering is considerably easier due to the higher
cross section for elastic events relative to inelastic processes) can distinguish between various predictions and provide valuable information regarding whether
the TPE contribution is sufficient to explain the form factor ratio discrepancy.

\subsection{The OLYMPUS Detector and Event Reconstruction}

\begin{figure}[thb!]
\centerline{\includegraphics[width=\columnwidth]{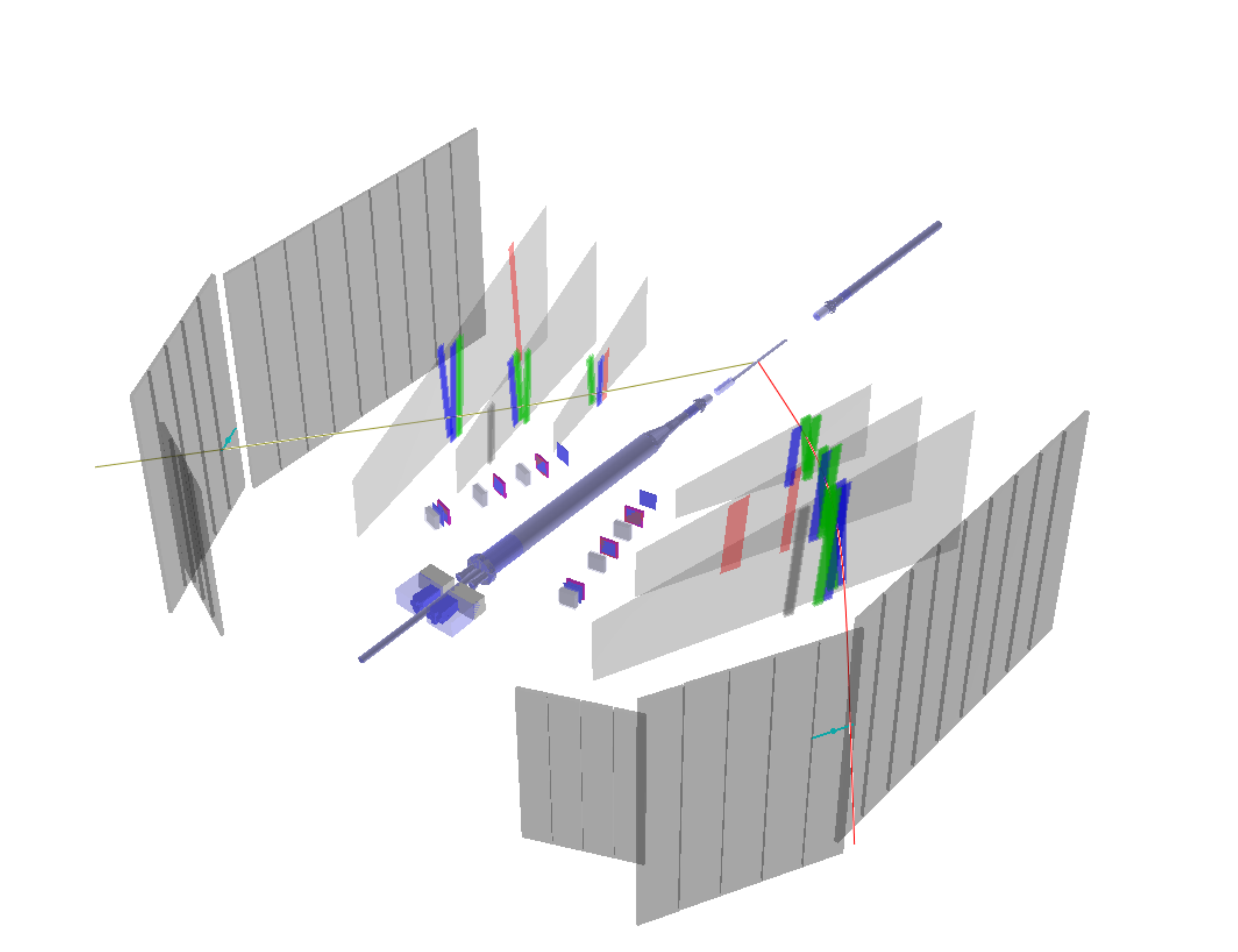}}
\caption{Reconstruction of an elastic \ep event from the OLYMPUS dataset, shown in the event display created for the experiment \cite{henderson}.  In the display,
the toroid coils and drift chamber frames are removed for clarity.}
\label{fig:event}
\end{figure}

The OLYMPUS experiment was conducted at the DORIS $e^\pm$ storage ring at DESY, Hamburg, Germany, with data taken over the course of two runs totaling three
months in 2012 and early 2013.  The experiment collected $\sim$4.4 fb$^{-1}$ of data, approximately equally split between the two lepton species.
Leptons of fixed energy (2.01 GeV) were incident on a hydrogen gas target internal to the DORIS ring \cite{Bernauer201420}.  The lepton species was switched daily to
minimize the effects of any longterm systematic changes in the beam or detector setup.  Elastic events
were exclusively reconstructed in a large acceptance spectrometer, with a toroidal magnetic field providing bending of particle trajectories for momentum reconstruction.
The detector consisted of the drift chambers and scintillator trigger panels previously used for the BLAST experiment at MIT-Bates
\cite{Hasell:2009zza} for the reconstruction of \pmp events, in combination with new detectors for monitoring of the luminosity.  Event reconstruction in the drift chambers
was conducted using an algorithm derived from the elastic arms algorithm \cite{OHLSSON1,OHLSSON2}, which is described in Reference \cite{russell}.
A schematic of the detector, along with a reconstructed elastic \ep event from the data is shown in Figure \ref{fig:event}.

\subsection{The OLYMPUS \pmp Analysis}

To measure the value of \ratio to high precision, an extremely detailed Monte Carlo simulation of the experiment was constructed.
This simulation accounted for the effects of detector acceptance, beam variations, and the magnetic field field that could lead to
false asymmetries between \ep and \pp events if not accounted for.  To develop this simulation, the physical configuration of the
detector and the magnetic field were extensively surveyed \cite{Bernauer20169}, and the various parameters of the experiment were
continuously monitored during data taking.  This information was used to generate a very high-statistics simulation dataset that 
matched the conditions of each experimental run.  The output of the simulation was convolved with measured detector efficiencies, resolutions,
etc. to produce output in the same format as the experimental data, allowing the reconstruction and analysis algorithms to be applied 
identically to each.  The value of \ratio was then reconstructed as
\begin{equation}
 R_{2\gamma} = \frac{\sigma_{e^+p}\left(Q^2\right)}{\sigma_{e^-p}\left(Q^2\right)} = \frac{N_{e^+,data}\left(Q^2\right)}{N_{e^-,data}\left(Q^2\right)}\cdot \frac{N_{e^-,MC}\left(Q^2,\mathcal{L}_{e^-}\right)}{N_{e^+,MC}\left(Q^2,\mathcal{L}_{e^+}\right)},
\end{equation}
where $N_{e^\pm}$ are the event counts in data and simulation of each species and the simulation results are weighted by the measured luminosities ($\mathcal{L}_{e^\pm}$)
of the data collected for each species.  Several independent event selection analyses were conducted using the reconstructed dataset to
provide cross checks and estimates of the systematic uncertainties due to analysis decisions for the final results \cite{henderson,schmidt,russell,ice,oconnor}.

This approach additionally conferred the advantage of allowing a simulation-based
approach to radiative corrections, in which the simulation events were associated with weights computed from a number of radiative
corrections models.  This allowed the full effect of acceptance constraints due to event selection to be accounted for, preventing
false asymmetries due to the differences in the \pp and \ep radiative corrections.  Details on this method may be found in References
\cite{schmidt} and \cite{russell} and will be the subject of an upcoming publication.

\subsection{Luminosity Monitoring}

As previously noted, the experiment utilized several methods of luminosity monitoring:
\begin{enumerate}
 \item measurement of the beam current and gas flow rate into the target, in combination
       with a molecular flow simulation of the target cell geometry \cite{henderson},
 \item elastic \pmp events at $\epsilon \approx 0.98$ ($\theta\approx12^\circ$) with the lepton detected in dedicated 6-plane tracking telescopes and
       the recoil proton in the upstream sections of the drift chambers \cite{henderson}, and
 \item monitoring of M{\o}ller, Bhabha, and elastic \pmp scattering events in a very forward ($\theta \approx 1.3^\circ$) PbF$_2$ symmetric calorimeter system
       \cite{PerezBenito20166}.
\end{enumerate}

\begin{figure}[thb!]
\centerline{\includegraphics[width=\columnwidth]{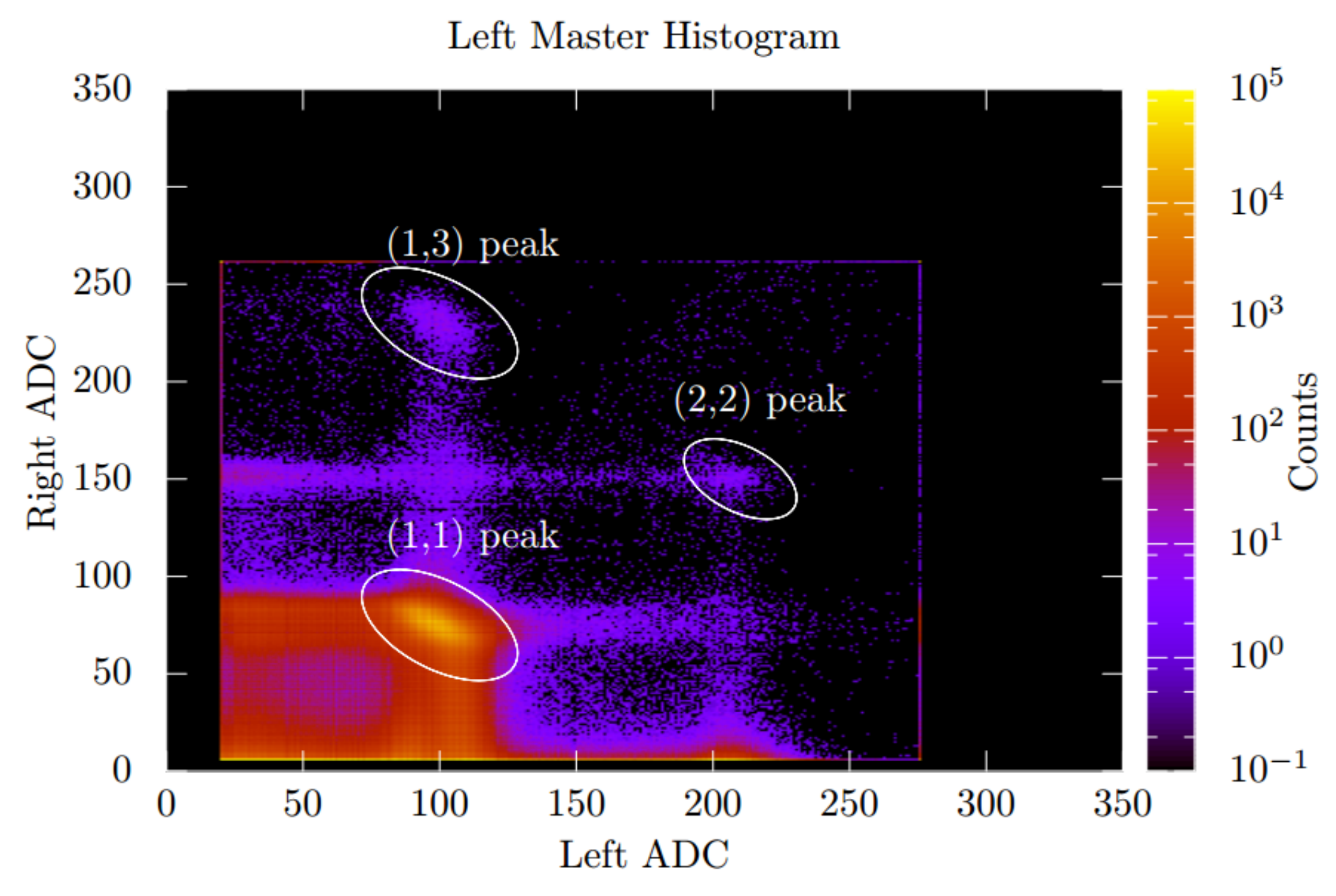}}
\caption{Two-dimensional histogram of the energy deposited in each side of the forward luminosity calorimeter, showing the contributions from the multi-interaction events (the
(1,3) peak), M{\o}ller/Bhabha scattering events (the (1,1) peak), and events in the (2,2) peak caused by double \pmp or double M{\o}ller/Bhabha events from the same bunch.
(Figure reproduced from \cite{schmidt}.)}
\label{fig:mie}
\end{figure}

The first method had the advantage of being completely independent of the physics of interest, but offered only $\sim$2\% relative, $\sim$5\% absolute
uncertainty in the luminosity measurement due to uncertainties in the temperatures of the target system components and the simulation used to determine
the effective target thickness.  The second method provided a high-statistics ($\sim$1\%/hour), well-determined measurement of the elastic \ep and \pp rates, but relied on the
assumption that the TPE contribution at the relevant kinematics is negligible.  The third method utilized multi-interaction events (MIE), in which an elastic \pmp event was
detected in coincidence with a Bhabha or M{\o}ller scattering event from the same beam bunch crossing.  Such an event resulted in the deposition of $\sim$3 GeV in one side of
the calorimeter from the combination of the very forward \pmp lepton and one of the M{\o}ller/Bhabha leptons and $\sim$1 GeV in the other from the remaining lepton.  By measuring
the rate of such events $N_{(1,3)}$ to the rate of M{\o}ller/Bhabha events for each lepton species $N_{(1,1)}$ (in which $\sim$1 GeV is deposited in each side of the calorimeter),
the luminosity could be extracted as:
\begin{equation}
 \mathcal{L}_\text{MIE} = \frac{N_{(1,3)}N_b}{N_{(1,1)}\sigma^\text{MC}_{e^\pm p}} + \text{Variance corrections},
\end{equation}
where $\sigma^\text{MC}_{e^\pm p}$ is the cross section for elastically-scattered leptons to reach the calorimeter as determined by simulation.  For reference, the variance
corrections (due to variations in the charge-per-bunch of the beam), amount to a $\sim$1\% correction on the measurement.  The energy deposition histogram, showing the contributions
from each of these event types is shown in Figure \ref{fig:mie}.  By measuring a ratio in this fashion, systematic
uncertainties due to detector efficiency and other effects are reduced and the uncertainty due to possible TPE contributions in the \pmp component of the measurement are 
much smaller than in the 12\dg measurement due to the very forward placement of the calorimeters. Complete details on this method may be found in 
Reference \cite{schmidt} and will be the subject of an upcoming publication.

The three luminosity monitors provided very consistent results, with the latter two methods each providing total statistical+systematic uncertainties of less than $0.5\%$ each,
well within the required precision for the \rtg measurement goals of the experiment.  Additionally, the independence of the 12\dg elastic reconstruction and the MIE luminosity
measurement allowed determination of the value of \rtg at an additional kinematic point outside the acceptance of the main spectrometer:
\begin{equation}
\begin{split}
 R_{2\gamma}\left(\epsilon = 0.98,Q^2 = 0.165\:\text{GeV}^2 \right) = \\ 0.9975 \pm 0.0010\:(\text{stat.}) \pm 0.0053\:(\text{syst.}).
\end{split}
\end{equation}
This measurement, in addition to providing a useful cross-check on the \rtg measurement in the main spectrometer acceptance, provides the most
precise existing determination of \ratio in the forward scattering region, providing discrimination between models at these kinematics and constraining
the forward elastic \pmp scattering luminosity normalization used for the VEPP-3 TPE experiment.  A future publication will detail this analysis, also
described in References \cite{schmidt} and \cite{henderson}.

\section{Summary and Upcoming Results}

\begin{figure}[tb!]
\centerline{\includegraphics[width=1.0\columnwidth]{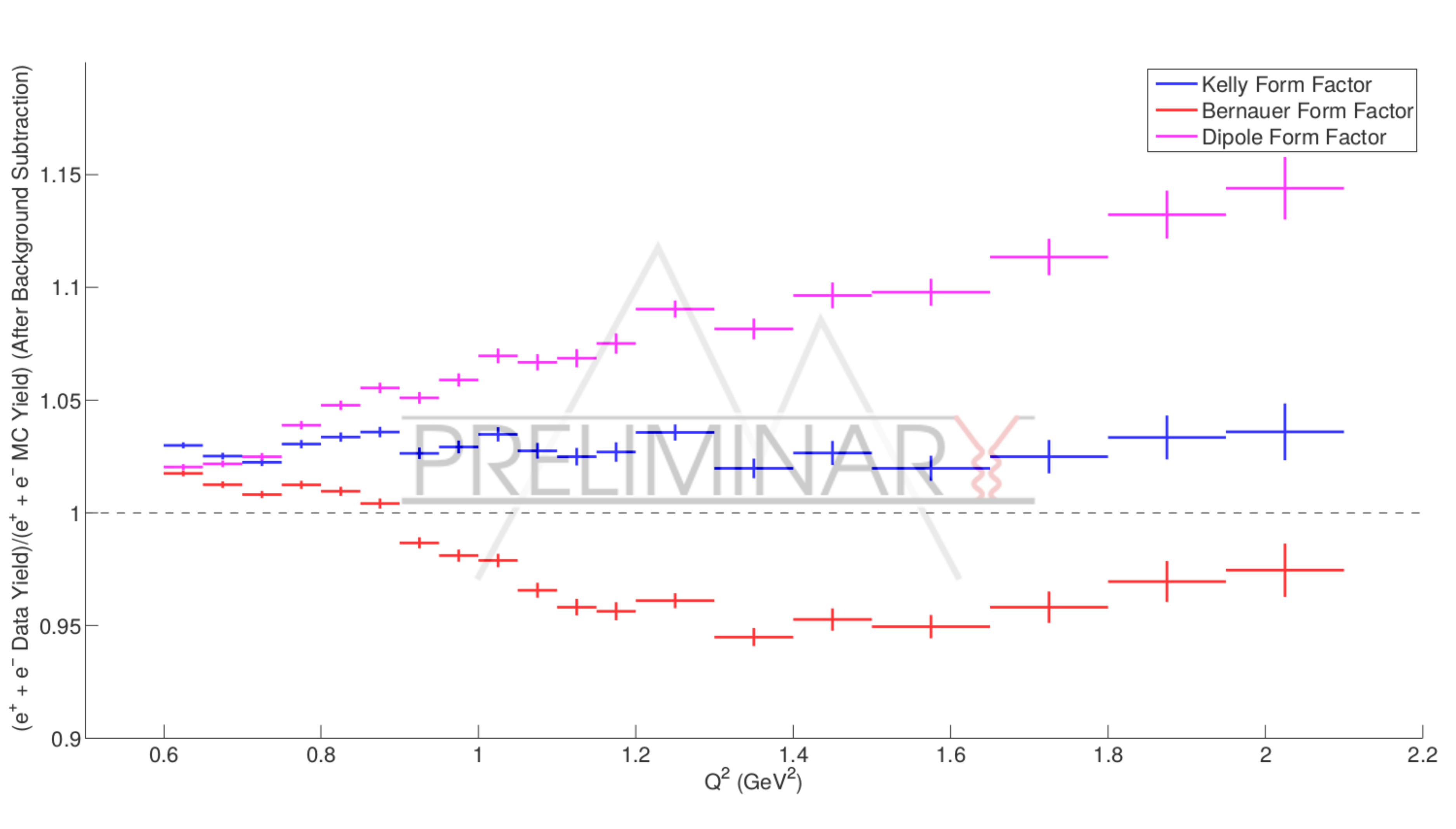}}
\caption{Ratio of the charge-summed ($e^+p+e^-p$) elastic scattering cross sections as determined by OLYMPUS to the prediction of
the simulation for several different form factor models \cite{KellyPhysRevC.70.068202,BerFFPhysRevC.90.015206}.  Note that this ratio
is normalized to the slow control luminosity only, and thus is subject to uncertainties in the absolute scale of $\sim$5\%.}
\label{fig:ffm}
\end{figure}

Since this presentation, the first results on \rtg from the OLYMPUS experiment have been released \cite{prlsubmit},
providing a high-precision determination of \ratio up to four-momentum transfers of $Q^2\approx 2.2$ (GeV/$c$)$^2$.
Critically, the OLYMPUS data provide an absolute normalization of the cross-section ratio using the MIE method
of luminosity determination and the tightest existing constraint on the value of \rtg at high-$\epsilon$ using
the 12\dg tracking system.  The results from the OLYMPUS experiment, in combination with those of the CLAS \cite{PhysRevLett.114.062003,clasprl} and VEPP-3 
\cite{vepp3PhysRevLett.114.062005} TPE experiments, constrain \rtg to within a few percent of unity for $Q^2\lesssim 2$ (GeV/$c$)$^2$.  While this is
a smaller TPE contribution to \pmp than predicted by some theoretical models, a definitive determination of whether TPE is responsible for the
entirety of the form factor ratio discrepancy at higher $Q^2$ will likely require experiments measuring \ratio at higher energies.  The initial OLYMPUS
publication on the \rtg result will be followed by several additional papers detailing the MIE luminosity determination, the high-$\epsilon$ \ratio
determination, radiative corrections for the OLYMPUS experiment, systematic uncertainties, and other topics.  In particular, the OLYMPUS data may also
provide constraints on proton elastic form factor models, as suggested by the discrimination power of the data between such models shown in Figure
\ref{fig:ffm}.

The combination of the results from the three TPE experiments places significant constraints on theoretical and phenomenological
descriptions of TPE.  While future measurements at higher $Q^2$ may be useful in determining the ultimate cause of the form
factor ratio discrepancy, significant progress has been made in exploring the problem below $Q^2\approx 2.2$ (GeV/$c$)$^2$
providing discrimination power between models used to predict behavior at higher energies.

\bibliography{references}

\end{document}